\documentclass[usegraphicx,useAMS,usenatbib]{mn2e}
\usepackage{amssymb}
\usepackage{aas_macros} 

\newcommand{\teff}{$T_{\mathrm{eff}}$}
\newcommand{\muhz}{$\mu$Hz}
\newcommand{\numax}{$\nu_{\mathrm{max}}$}
\newcommand{\nuac}{$\nu_{\mathrm{ac}}$}
\newcommand{\dnu}{$\Delta\nu$}
\newcommand{\astec}{\textsc{astec}}
\newcommand{\adipls}{\textsc{adipls}}
\newcommand{\yrec}{\textsc{yrec}}

\title[The relation between \dnu\ and \numax]{The relation between \dnu\ and \numax\ for solar-like oscillations}   
\author[D. Stello et al.]
{D. Stello$^{1}$,%\thanks{E-mail: stello@physics.usyd.edu.au (DS)} 
 W.~J. Chaplin$^{2}$,
 S. Basu$^{3}$, 
 Y. Elsworth$^{2}$ and 
 T.~R. Bedding$^{1}$\\ 
$^{1}$Sydney Institute for Astronomy (SIfA), School of Physics,
  University of Sydney, NSW 2006, Australia\\ 
${2}$School of Physics and Astronomy, University of Birmingham,
  Edgbaston, Birmingham, B15 2TT, UK\\
$^{3}$Department of Astronomy, Yale University, P.O. Box 208101, New Haven, CT 06520-8101, USA}
\begin{document}
\date{Accepted 2009 September 28.  Received 2009 September 27; in original form
2009 September 3}
\pagerange{\pageref{firstpage}--\pageref{lastpage}} \pubyear{2009}
\maketitle
\label{firstpage}
\begin{abstract}
Establishing relations between global stellar parameters and asteroseismic
quantities can help improve our understanding of
stellar astrophysics and facilitate the interpretation of observations.
We present an
observed relation between the large frequency separation, \dnu, and the
frequency of maximum power, \numax. We 
find that \dnu\ $\propto \nu_{\mathrm{max}}^{0.77}$, allowing
prediction of \dnu\ to about 15 per cent given \numax. 
Our result is further supported by established scaling relations for \dnu\ and
\numax\ and by extended stellar model calculations, which confirm
that \dnu\ can be estimated using this relation for basically any star
showing solar-like oscillations in the investigated range ($0.5<M/$M$_\odot<4.0$).  
%%AandA
%\keywords{stars: fundamental parameters --- stars: oscillations --- stars:
%  interiors}
%}
%\maketitle
%%MNRAS
\end{abstract}

\section{Introduction} 
%The Sun has shown us some remarkable results of physics. 
%It was discovered
%by \citet{Leighton62} through time-series observations of the Sun's radial velocity
%that the Sun showed periodic variation of roughly 5 min, which was 
%later interpreted as global oscillations by
%\citet{Ulrich70,LeibacherStein71}.  The oscillations were subsequently 
%detected as brightness fluctuations as well \citep{WoodardHudson83}. 

Simple scaling relations for asteroseismic quantities have proven very
useful when analysing stellar oscillations.  In particular, relations for
scaling the quantities \numax\ \citep{Brown91} and \dnu\ \citep{Ulrich86}
from the Sun have been widely used to constrain stellar global parameters,
or simply to verify the asteroseismic signal of stars that already had 
well-constrained stellar parameters. Here, \numax\ is the frequency of
maximum power of the oscillations, and \dnu\ is the sol-called large
frequency separation between consecutive overtones. 
For the Sun, \numax\ $\simeq3100\,$\muhz\ and \dnu\ $\simeq135\,$\muhz.
%The characteristic signature of the Solar five-minute oscillations is a
%large number of roughly equally spaced peaks, like a picket fence, in the
%Fourier transform of a whole disk integrated velocity or intensity time
%series.  Each peak correspond to a single mode of oscillation and their
%heights %, which is related to the mode amplitudes, 
%are modulated by a broad
%envelope controlled by the excitation and damping processes.  
%For the Sun, the maximum height of the envelope is located at \numax\
%$\simeq3100\,$\muhz\ and the characteristic frequency separation between modes
%of consecutive radial order $n$ is \dnu\ $\simeq135\,$\muhz. 
%if start with a punchy sentence, rewrite entire intro
%The Solar five-minute oscillations are caused by damped standing sound waves from
%thousands of eigenmodes that get continuously re-excited by near-surface
%convection.  We see this as a large number of roughly equally spaced peaks,
%like a picket fence, in the Fourier transform of a whole disk integrated
%velocity or photometric time series.  Each peak correspond to a single mode 
%of oscillation and their heights, which is related to the mode amplitudes,
%are modulated by a broad envelope controlled by the excitation and damping
%processes.  
%For the Sun, the maximum amplitude of the envelope is located at \numax\
%$\simeq3100\,$\muhz\ and the characteristic frequency separation between modes
%of consecutive radial order $n$ is \dnu\ $\simeq135\,$\muhz. 

%[MAYBE IN INTRO MENTION ASYMPTOT THEORY AND KB95 SCALING RELATIONS AND
%JUSTIFY THAT DNU AND NUMAX CAN CONSTRAIN STELLAR PARM?]
Over the past decade, detections of solar-like oscillations in other
stars have been growing in number \citep[see reviews
  by][]{Aerts08,BeddingKjeldsen08}, and the \textit{CoRoT} space mission has caused
a recent surge \citep{Michel08,Ridder09}.  This flow of data is expected
to increase to a flood with the \textit{Kepler} mission \citep{Dalsgaard08}.  %Detailed
				%analysis of the individual mode
				%frequencies 
%have led to exciting new insights to the Sun and the physical processes
%within \citep{Dalsgaard02}. However, 
For many stars, knowing \dnu\ or \numax\ can help constrain the stellar
parameters considerably \citep{Stello08,Kallinger08a,Stello09,Miglio09} and
hence yield important new insights into stellar structure and evolution, even
without a full asteroseismic analysis. 
In addition, recent developments of automated software for asteroseismic
data analysis benefit from knowing a simple relation between \dnu\ and
\numax\ \citep{Mathur09,Huber09,Hekker09}.

In this Letter we investigate the relation between \dnu\ and \numax,
building on previous studies to show how these two quantities are related
to each other and to the global stellar parameters. We also carry out
stellar model calculations to support this investigation, including a
comparison of stellar models with the established scaling relations for \dnu\
and \numax.
%with the aim of supporting developments of automated software for asteroseismic
%data analysis.

\section{The observed relation}\label{observations}
\begin{figure}
\includegraphics[width=84mm]{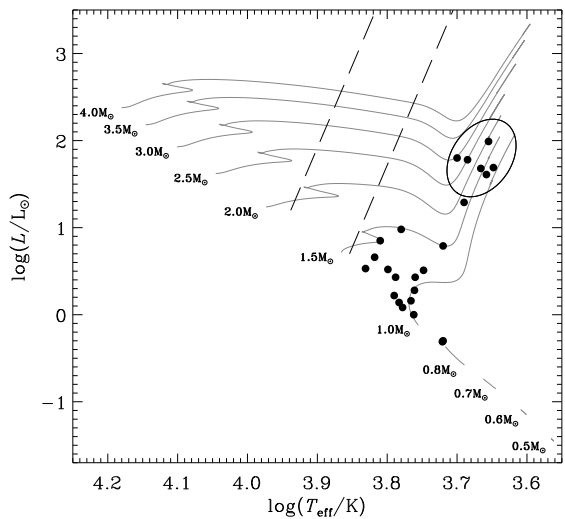}
%\epsscale{1.0}
%\plotone{f0.eps}
\caption{$H$--$R$ diagram showing all stars listed in Table~\ref{tab1}. The
  ellipse indicates the approximate location of the 31 \textit{CoRoT} red
  giants. The evolutionary tracks (grey curves), illustrate the 
  range in mass and evolutionary state that we investigate with our stellar
  models (see Sect.~\ref{models}). The
  dashed lines show the approximate location of the classical instability strip
  \citep{SaioGautschy98}. 
\label{f0}} 
\end{figure}

Table~\ref{tab1} shows published measurements of \dnu\ and \numax\ for 55
stars 
%observed in radial velocity, 
including the Sun. 
%This list comprises mostly main
%sequence and subgiant stars, but also contains three red giants.  We also include
%photometric results, mostly on red giants, from the \textit{WIRE} and \textit{CoRoT} space missions.
%Unfortunately, it is still not common practise to publish a measurement of
%\numax, and we would like to encourage people to do so. 
In some cases, where the value of \numax\ was not specified, we have
estimated it from the published power spectra.  The location of the stars in the
$H$--$R$ diagram are shown in Fig.~\ref{f0}. 

\begin{table}
\begin{center}
\caption{Published measurements of \dnu\ and \numax \label{tab1}}
\begin{tabular}{lrrl}
\hline
Star            & \dnu\   & \numax\ & Source \\
                & (\muhz) & (\muhz) &        \\
\hline
$\tau\,$Cet     & 170    & 4500    & \citet{Teixeira09}   \\
$\alpha\,$Cen B & 161.4  & 4100    & \citet{Kjeldsen07}   \\
Sun             & 134.8  & 3100    & \citet{Kjeldsen07}   \\
$\iota\,$Hor    & 120    & 2700    & \citet{Vauclair08}   \\
$\gamma\,$Pav   & 120.3  & 2600    & \citet{Mosser08}     \\
$\alpha\,$Cen A & 106.2  & 2400    & \citet{Kjeldsen07}    \\
HD175726        &  97    & 2000    & \citet{Garcia09}     \\
$\mu\,$Ara      &  90    & 2000    & \citet{Bouchy05}     \\
HD181906        &  87.5  & 1900    & \citet{Garcia09}     \\
HD49933         &  85.9  & 1760    & \citet{Garcia09}     \\
HD181420        &  75    & 1500    & \citet{Garcia09}     \\
$\beta\,$Vir    &  72    & 1400    & \citet{Carrier05a}   \\
$\mu\,$Her      &  56.5  & 1200    & \citet{Bonanno08}    \\
$\beta\,$Hyi    &  57.5  & 1000    & \citet{Kjeldsen07}    \\
Procyon         &  55    & 1000    & \citet{Arentoft08}     \\
$\eta\,$Boo     &  39.9  &  750    & \citet{Carrier05}    \\
$\nu\,$Ind      &  25.1  &  320    & \citet{Kjeldsen07}    \\
$\eta\,$Ser     &   7.7  &  130    & \citet{Barban04}     \\
$\xi\,$Hya      &   6.8  &   90    & \citet{Frandsen02}   \\
$\beta\,$ Vol   &   4.9  &   51    & unpublished \textit{WIRE} data  \\
$\epsilon\,$Oph &   5.3  &   50    & \citet{Barban07}          \\
$\xi\,$ Dra     &   4.0  &   36    & unpublished \textit{WIRE} data  \\
$\kappa\,$ Oph  &   4.5  &   35    & unpublished \textit{WIRE} data  \\
HR3280          &   3.2  &   25    & unpublished \textit{WIRE} data  \\
31 \textit{CoRoT} giants &   2--7 &  15--73 & \citet{Kallinger08a}    \\
\hline
\end{tabular}
\end{center}
\end{table}

\begin{figure}
\includegraphics[width=84mm]{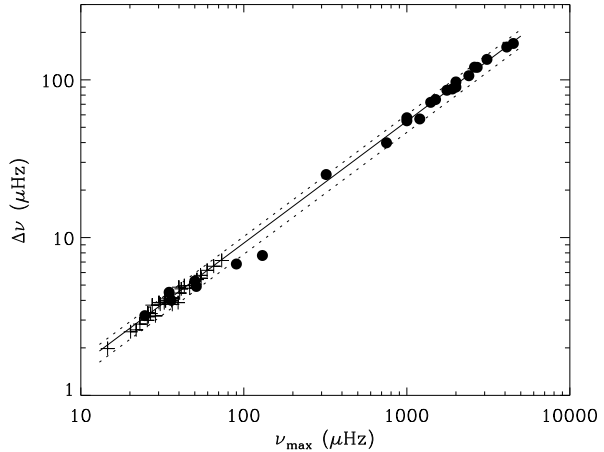}
%\epsscale{1.0}
%\plotone{f1.eps}
\caption{Observed \dnu\ versus \numax\ for the stars listed in
  Table~\ref{tab1}. The 31 \textit{CoRoT} red giants are plotted with plus
  symbols. The solid line is a power-law fit, and dotted lines show $+10$ per cent
  and $-15$ per cent deviations. 
\label{f1}} 
\end{figure}
%notable,noteworthy,noticeable,remarkable,striking,surprising
In Fig.~\ref{f1} we plot \dnu\ versus \numax\ for the data listed in
Table ~\ref{tab1}. We see a remarkably tight relation over nearly three
orders of magnitude. A power-law fit gives %5 per cent if ignore two
				%outliers  
\begin{eqnarray}
\Delta\nu=(0.263\pm0.009)\mu\mathrm{Hz}\, (\nu_{\mathrm{max}}/\mu\mathrm{Hz})^{0.772\pm0.005}. 
\label{obsrel}
\end{eqnarray}
The two stars that deviate the most from this relation are $\eta\,$Ser and
$\xi\,$Hya.  In both cases, there are reported ambiguities in the
determination of the large separation, which could explain the deviation. 
%[IF LETTER FORMAT REMOVE THE REST OF THIS PARAGRAPH OR AT
%  LEAST THE LAST TWO SENTENCES] 
%It is noteworthy that separations of 11.4\muhz\ and 8.2\muhz\ 
%for $\eta\,$Ser and $\xi\,$Hya, respectively, would follow our observed
%relation much more closely, which in both cases correspond to observed
%peaks in the published autocorrelation functions. 
Apart from those two cases, all stars in Fig.~\ref{f1} fall within $+10$ per cent
and $-15$ per cent of the fitted relation, while the main sequence stars alone
fall within $\pm5$ per cent. 
We note that \citet{Hekker09} have confirmed the tight correlation
for a sample of several hundred red giants observed by \textit{CoRoT}, which also
seems to indicate a tendency for some stars to fall below the relation. 
%In addition, both
%datasets were obtained with the \textsc{Coralie} 
%spectrograph and reduced using the optimum-weight procedure
%\citep{Bouchy01}, which indirectly enforces high-pass filtering, hence
%potentially shifting the measured \numax\ to a higher frequency. While
%$\eta\,$Ser is probably not affected significantly, it seems a likely
%contributor for $\xi\,$Hya \citep{Frandsen02}.

%Despite the relatively broad range of stars in our sample, we would like
%to test the relation with models and use a larger region of the $H$--$R$
%diagram (see Fig.~\ref{f0}). Using dense model grids will further allow
%us to investigate the dependencies on the global stellar parameters. 

%In the following, we will introduce/describe the stellar models before we
%turn to their use in the subsequent sections for the further discussion and
%interpretation/understanding of this observed relation.

\section{Scaling relations}\label{scaling}
\begin{figure}
\includegraphics[width=84mm]{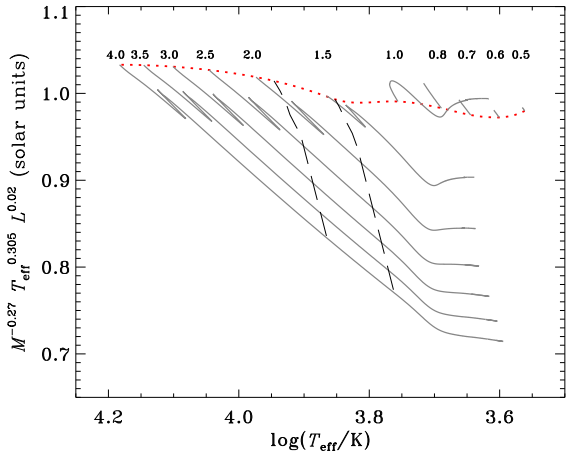}
%\epsscale{1.0}
%\plotone{f8.eps}
\caption{The scaling factor in square brackets in Eq.~(\ref{dnunumaxrel})
  for $a=0.77$, for the same evolutionary tracks as in Fig.~\ref{f0}.
  The horizontal dotted curve is the ZAMS and the dashed lines indicate the instability
  strip. Masses are indicated in solar units.
\label{f8}} 
\end{figure}

%In the following we discuss how the observed relation between \dnu\ and \numax\ 
%(see Fig.~\ref{f1}) is linked to previously published scaling
%relations of those two quantities.
Can previously published scaling relations for
\dnu\ and \numax\ explain the observed relation between
them?  It is well-established \citep[e.g.][]{Ulrich86} that,
to a good approximation, \dnu\ is proportional to the square root of the
stellar density:  
%$\Delta\nu=\sqrt{\rho/\rho_\odot}\Delta\nu_\odot$
\begin{eqnarray}
\frac{\Delta\nu}{\Delta\nu_{\odot}}=\sqrt{\frac{\rho}{\rho_\odot}}%\nonumber\\
                     =\frac{(M/M_\odot)^{0.5}(T_{\mathrm{eff}}/T_{\mathrm{eff,\odot}})^3}{(L/L_\odot)^{0.75}}.
\label{dnusc}
\end{eqnarray}
Also, following \citet{Brown91} and \citet{KjeldsenBedding95}, we expect \numax\
to scale as the acoustic cut-off frequency, \nuac. Hence,
\begin{eqnarray}
\frac{\nu_{\mathrm{max}}}{\nu_{\mathrm{max,\odot}}}=\frac{\nu_{\mathrm{ac}}}{\nu_{\mathrm{ac,\odot}}}%\nonumber\\
                  =\frac{M/M_\odot (T_{\mathrm{eff}}/T_{\mathrm{eff,\odot}})^{3.5}}{L/L_\odot}, 
\label{numaxsc}
\end{eqnarray}
where it is observed for the Sun that $\nu_{\mathrm{ac,\odot}} \simeq
1.7\nu_{\mathrm{max,\odot}}$ \citep{BalmforthGough90,Fossat92}.
%\begin{eqnarray}
%\nu_{\mathrm{ac,\odot}} \simeq 1.7\nu_{\mathrm{max,\odot}}\,.
%\label{numaxnuacsol}
%\end{eqnarray}
Given that \dnu\ and \numax\ scale differently with stellar parameters, the
tightness of the correlation in Fig.~\ref{f1} is
perhaps surprising.  To understand this, we raise Eq.~(\ref{numaxsc})
to the power $a$ and divide by Eq.~(\ref{dnusc}) which, after
rearranging, gives 
\begin{eqnarray}
\frac{\Delta\nu_{\mathrm{}}}{\Delta\nu_{\mathrm{\odot}}}=\left[\frac{(M/M_\odot)^{0.5-a}(T_{\mathrm{eff}}/T_{\mathrm{eff,\odot}})^{3-3.5a}}{(L/L_\odot)^{0.75-a}}\right]\left(\frac{\nu_{\mathrm{max}}}{\nu_{\mathrm{max,\odot}}}\right)^{a}. 
%\dnu\
%$=M^{0.5-a}T_{\mathrm{eff}}^{3-3.5a}L^{a-0.75}\nu_{\mathrm{max}}^a$ 
\label{dnunumaxrel}
\end{eqnarray}
The tight relation in Fig.~\ref{f1} leads us to conclude that the scaling
factor in square brackets in Eq.~(\ref{dnunumaxrel}) must be approximately
constant when $a=0.77$. Indeed, this is easy to see because that value of
$a$ eliminates almost completely the parameter that varies the most over the
stars being considered, namely $L$, and leaves a very weak dependence 
on $M$ and \teff. %, and therfore a combined variation of $L$, $M$, and \teff\
%that is much smaller than the orders of magnitude variation in \numax
%across the same range of stars. %P\equiv 
This is illustrated in Fig.~\ref{f8}, which shows the scaling factor 
%$M^{0.5-a}T_{\mathrm{eff}}^{3-3.5a}L^{a-0.75}$ 
for the observed value $a=0.77$. % is practicly constant compared to the orders of magnitude
%variation in \numax\ across the same range of models. 
The scaling factor indeed covers a band 25 per cent wide for the region sampled by
the stars (see Fig.~\ref{f0}), in agreement with the $+10$ to
$-15$ per cent deviation shown by the observational data in Fig.~\ref{f1}. 
Figure~\ref{f8} further supports the lower scatter observed for the main
sequence stars and the higher, slightly skewed, scatter for red giants.
%With a variation of less than 30 per cent it is
%practically constant campared to the orders of magnitude
%variation in \numax\ across the same range of models.
%, which clearly
%illustrates it's small variation campared to the orders of magnitude
%variation in \numax\ across the same range of models.
%To obtain the strongest correlation between \dnu\
%and \numax\ one have to minimise the variation of the scaling factor %the quantity
%$P\equiv M^{0.5-a}T_{\mathrm{eff}}^{3-3.5a}L^{a-0.75}$ 
%by adjusting $a$.  
We have determined
the value of $a$ that minimizes the spread in the scaling factor and found
that it %$P$ %this quantity 
depends slightly on the stellar parameter range that we consider.  For example,
consideration of the entire model grid requires $a=0.78$, %resulting in less
%than 40 per cent variation in $P$, 
while considering only ZAMS models gives $a=0.75$. %, which then results in less

\section{Models}\label{models}
We now expand our investigation of the \dnu--\numax\
relation using two sets of stellar
models. One is a dense grid, comprising over a million models derived
using the Aarhus stellar evolution code \astec\ \citep{DalsgaardAstec08}
and the adiabatic pulsation code \adipls\ \citep{DalsgaardAdipls08}.  
This grid is slightly expanded in parameter space but otherwise identical
to the one constructed by \citet{Stello09} (see references herein). 
%We refer to \citet{Stello09} for further details on the construction of
%this grid. 
The other is a set derived using the Yale code \yrec\ \citep{Demarque08}
with mixing-length parameter $\alpha=1.80$, initial hydrogen
abundance $X=0.707$, and metallicity $Z=0.018$.
Although the set of \yrec\ models is more sparse, with a total 
of 92 models, it spans almost uniformly the same range in parameter space
as the \astec\ grid.  
The region covered by the models is indicated by the evolutionary tracks in
Fig.~\ref{f0}.  

%The \astec\ models were generated using the EFF equation of state
%\citep{Eggleton73}, a fixed mixing-length parameter, $\alpha=1.8$, and an
%initial hydrogen abundance of $X=0.7$. The opacities were calculated using
%the solar mixture of \citet{GrevesseNoels93} and the opacity tables of
%\citet{RogersIglesias95} and \citet{Kurucz91} ($T<10\,000$\,K). 
%Rotation, overshooting and diffusion were not included.
%The range and resolution of the parameters in the grid
%is $0.5<M<2.3$ $\Delta M=0.01$, $0.009<Z<0.55$ $\Delta \log(Z)=0.1$, and
%the models were evolved to the base of the RGB. %or more if use RGB grid extent
%
%The \yrec\ models were constructed with the OPAL equation of state (Rogers et
%al. 1996), OPAL high temperature opacities (Rogers \& Iglesias 1995;
%Iglesias \& Rogers 1996) and Alexander \& Ferguson (1994) low temperature
%opacities. The nuclear reaction rates from Adelberger et al. (1998) but
%with the $^{14}N(p,\gamma)^{15}O$ rate of Formicola et al. (2004). The
%modes were construted with $\alpha=1.80$. Opacities were calculated assuming
%the heavy-element mixture of Grevesse \& Noels (1993).
%Diffusion was not included.

In addition to the global stellar parameters $M$, $L$, and \teff, we
calculated for each model the large frequency separation as the inverse of  %in three
				%different ways: from scaling the solar
				%value, 
%\begin{equation}
%\Delta\nu_{\mathrm{sc}}=(M/R^3)^{0.5}\Delta\nu_\odot; 
%\label{dnusc}
%\end{equation}
the sound travel time through the star: 
\begin{equation}
\Delta\nu_{\mathrm{}}=\Big[2\int_0^Rc^{-1}dr\Big]^{-1},
\label{dnuint}
\end{equation}
where $c$ is the sound speed \citep{Tassoul80,Gough86}.
%and from the slope of a first order polynomial fit to 11 consecutive radial
%orders around \numax$_{\mathrm{sc}}$,
%\begin{equation}
%\Delta\nu_{\mathrm{fit}}=\frac{\sum_{n=n_{\mathrm{max}}-5}^{n_{\mathrm{max}}+5}nf_n}
%                              {\sum_{n=n_{\mathrm{max}}-5}^{n_{\mathrm{max}}+5}n^2}. 
%\label{dnufit}
%\end{equation}
%We further confirmed that a
%fourth method, using the area of the trapezoid defined by the consecutive
%radial modes gave basically the same result as \dnu$_{\mathrm{fit}}$.
We also calculated the acoustic cutoff frequency by %in two/three different ways: from scaling the
%solar value, 
%\begin{equation}
%\nu_{\mathrm{max,sc}}=M/(R^2T_{\mathrm{eff}}^{0.5})\nu_{\mathrm{max,\odot}}; 
%\label{nuasc}
%\end{equation}
%and 
assuming an isothermal atmosphere, which gives \citep{BalmforthGough90}
\begin{equation}
\nu_{\mathrm{ac}}=\frac{c}{2H}, 
\label{nuaciso}
\end{equation}
evaluated at the surface ($T=$ \teff). Here, %$c$ is the sound speed and 
$H=p/(g\rho)$ is the density scale height, $p$ is pressure, $g$ is 
gravity, and $\rho$ is density. % and use \numax\ $=$ \nuac/1.8 \citep{Brown91}.  
%[HOW MUCH SHOULD WE GO IN DETAIL WITH THESE VARIOUS WAYS OF CALULATING DNU
%AND NUAC? I DID NOT MENTION THE TRAPEZ VERSION OF DNU, SINCE TO MY
%KNOWLEDGE NOBODY IS REALLY FAMILIAR WITH THIS APPROACH AND THE RESULTS ARE
%ALMOST IDENTICAL WITH THE MORE COMMEMLY USED FIT TO CONSECUTIVE RADIAL ORDER
%MODES. IN OTHER WORD IT DOES NOT ADD MUCH NEW, SO NO REASON TO DILUTE THE
%DISCUSSION WITH IT...OBJECTIONS? ]
\begin{figure}
\includegraphics[width=84mm]{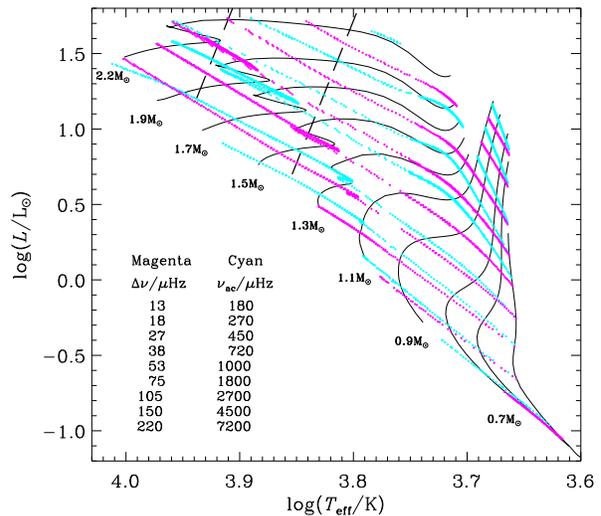}
%\epsscale{1.0}
%\plotone{f2.eps}
\caption{$H$--$R$ diagram of grid subset (\astec\ models with
  $Z=0.014$). Models within a narrow range of fixed values of
  $[2\int dr/c]^{-1}$ (magenta) and $c/2H$ (cyan) are
  indicated. The dashed lines indicate the instability 
  strip. Note that the low-mass models have been evolved beyond the
  age of the universe.
\label{f2}} 
\end{figure}

In Fig.~\ref{f2} we show a subset of the \astec\ grid for a fixed
metallicity ($Z=0.014$).  Contours of constant $[2\int dr/c]^{-1}$
(magenta) are almost parallel to contours of constant $c/2H$ (cyan),
indicating a 
strong correlation.  Hence, knowing one of these two quantities gives a
good estimate of the other.  This is particularly pronounced in the
lower-right corner of the diagram, corresponding to cool main-sequence and 
subgiant stars. % with masses below $\sim$1.1 M$_\odot$.  
A somewhat weaker correlation is seen in the red giant phase.
%, the two sets of contours are less
%parallel in the red giant phase, indicating a somewhat weaker correlation.
Note that we
neglected the post He-core burning phase, including any mass loss
associated with red giant branch evolution. If included, it would
blur the contours in this region \citep{Stello08}.  
%However, in reality the tight relation in Fig.~\ref{f1}
%suggests that this information is of rather limited use.
%The plot has been extended with low mass models going past the age of the universe
%for better visual impression of the trends/relation between \dnu\ and \numax\
%for cool end of the diagram.
%\begin{figure}
%\includegraphics[width=84mm]{f7.eps}
%%\epsscale{1.0}
%%\plotone{f7.eps}
%\caption{\dnu\ versus \nuac\ relation for the \yrec\ models. Blue symbols are
%  models hotter than the red edge of the instability strip, while red and grey
%  symbols are cooler. Red symbols are low mass ($M<1.2$M$_\odot$). Solid
%  line is a power law fit and dotted lines show $+10$ per cent and $-15$ per cent deviation.  
%\label{f7}} 
%\end{figure}
\begin{figure}
\includegraphics[width=84mm]{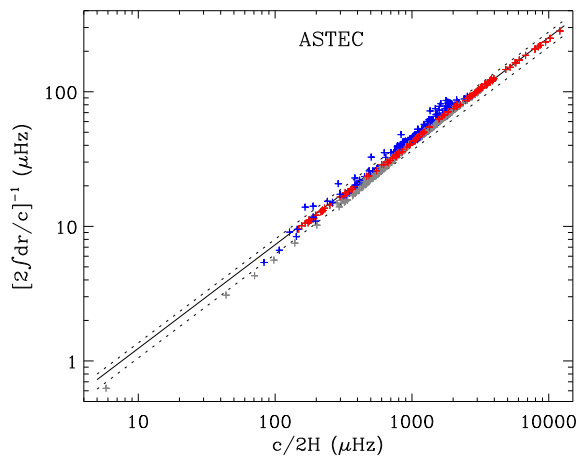}
%\epsscale{1.0}
%\plotone{f6.eps}
\caption{\dnu\ versus \nuac\ relation for the \astec\ models. Blue symbols are
  models hotter than the red edge of the instability strip, while red and grey
  symbols are cooler. Red symbols are low mass ($M<1.2$M$_\odot$). Solid
  line is a power-law fit and dotted lines show $+10$ per cent and $-15$
  per cent deviations.
%Same as Fig.~\ref{f7} but for a subset 
%of the \astec\ grid. 
\label{f6}} 
\end{figure}

Figure~\ref{f6} shows the relation between \dnu\ and \nuac\
for the \astec\ models.  We find very similar results for the \yrec\
models.  Stars hotter than 
the red edge of the instability strip are generally not expected to exhibit
solar-like oscillations and are shown in blue, while cool models
are shown in red and grey.
%were further divided in to low (red) and high (grey) mass, which seems to show
%two distinct populations especially for the dense \astec\ grid.
We compare the models with observations by fitting a power law with the observed
value of $a=0.77$ fixed.  We see
excellent agreement with observations, particularly at low masses
($<1.2\,$M$_\odot$, red), which is perhaps not surprising since the
observations are predominantly from low-mass stars. 
Interestingly, all cool main-sequence models -- high values of \nuac\ --
show a very tight relation, while for more evolved models -- lower values
of \nuac\ -- the higher mass ones (grey) 
scatter more and fall below the power-law relation. %with an indication of a negative skew. 
This agrees with the
observations and with the scaling factor in Eq.~(\ref{dnunumaxrel})
(see also Fig.~\ref{f8}).
Finally, we found that the theoretical \dnu--\nuac\ relation was not notably
sensitive to metallicity.
However, it remains to be explained why \dnu, which depends on the stellar
mean density, correlates so strongly with \nuac, which depends on the local
conditions in the atmosphere.
%Figures~\ref{f7} and \ref{f6} show the relation between $[2\int dr/c]^{-1}$ and
%$c/2H$ for the \yrec\ and \astec\ models, respectively.  Because stars hotter than 
%the red edge of the instability strip are generally not expected to show
%solar-like oscillations we divided the models into hot (blue) and cool.  The
%cool models were further divided into low (red) and high (grey) mass.  

\section{Discussion}\label{discussion}
Several points should be kept in mind when comparing the model calculations with the
observations.  Firstly, we do not measure \nuac\ directly in stars and
instead rely on its relation to \numax\ (see
Eq.~\ref{numaxsc}), assuming that \numax/\nuac\ does not 
vary from the solar value.  In addition, \numax\ can be difficult to define
and measure, especially for stars with broad or double-humped envelopes, as
seen in Procyon \citep{BeddingKjeldsen06b,Arentoft08}. 
On the other hand, \dnu\ varies with frequency and mode degree, and its
measurement from the power spectrum gives an average that will not be
exactly equal to $[2\int dr/c]^{-1}$. %However, the agreement will
%usually be within a few per cent.
That said, Fig.~\ref{f6} clearly supports the
tight \dnu--\numax\ relation that we observe.

\begin{figure}
\includegraphics[width=84mm]{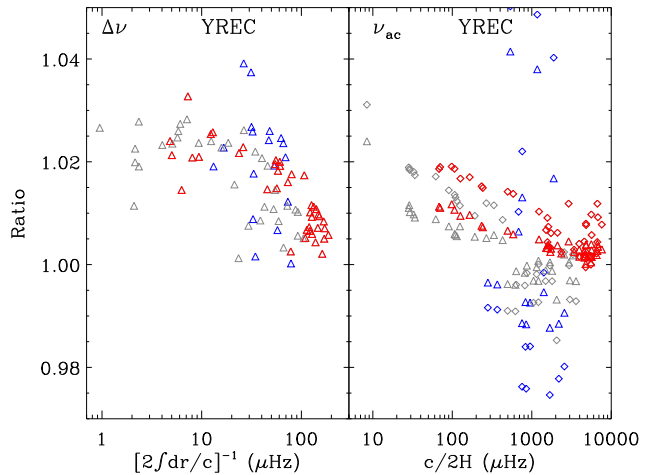}
%\epsscale{1.0}
%\plotone{f10.eps}
\caption{Left panel: Ratio between $[2\int dr/c]^{-1}$ and a
  linear fit to model frequencies, \dnu$_{\mathrm{fit}}$.
  Right panel: Ratio between $c/2H_2$ and $c/2H_1$ where $H_2=-dr/d\ln
  \rho$ and $H_1=p/(g\rho)$ (triangles), and between
  $(c/2H_1)\sqrt{1-2dH/dr}$ and $c/2H_1$ (diamonds). Both panels show
  results for \yrec\ models. Colour notation follows that of
  Fig.~\ref{f6}. 
\label{f10}} 
\end{figure}

The preceding comments relate to measuring \dnu\ and \numax\ from
observations. We should also consider the various ways in which these
quantities can be estimated from the models.  So far we have used
Eqs.~(\ref{dnuint}) and~(\ref{nuaciso}).  Here, we derive the
large frequency separation and the acoustic cutoff frequency from the 
models in other ways and investigate any significant systematic differences.  
The large separation was derived by fitting to 11 consecutive radial
orders around \numax, which we denote
\dnu$_{\mathrm{fit}}$. The result for the \yrec\ models (Fig.~\ref{f10},
left panel) shows 
agreement within a few per cent between $[2\int dr/c]^{-1}$ and
\dnu$_{\mathrm{fit}}$, although we note that the ratio $[2\int
  dr/c]^{-1}/$\dnu$_{\mathrm{fit}}$ is always higher than unity. 
We see similar results for the \astec\ models.
%\begin{figure}
%\includegraphics[width=84mm]{f5.eps}
%%\epsscale{1.0}
%%\plotone{f5.eps}
%\caption{Ratio between $[2\int dr/c]^{-1}$ and a
%  linear fit to model frequencies, \dnu$_{\mathrm{fit}}$, for the \yrec\
%  models. Colour notation follows that of Fig.~\ref{f6}.     
%\label{f5}} 
%\end{figure}

We then tested two additional ways to calculate the acoustic cutoff
frequency. Firstly, $H$ was derived from the actual density gradient in
the model, but still using Eq.~(\ref{nuaciso}). Secondly 
we used the full expression $\nu_{\mathrm{ac}}=(c/2H)\sqrt{1-2dH/dr}$
\citep{BalmforthGough90}. 
The results are shown in Fig.~\ref{f10} (right panel). Apart from a few hot models that
deviate by up to 25 per cent, the agreement is again within a few per cent. 
For our purpose of investigating the relation between
\dnu\ and \nuac, a few per cent difference in either quantity is not important, but they should
be kept in mind for other applications. %[ANY REASON TO COMMENT ON TRENDS
%  FOR DIFF GROUPS OF MODELS (COLOURS)?] %In the
%following, we adopt $(2\int dr/c)^{-1}$ as our model value of \dnu, %because
%it varies more smoothly for evolved stars, which have a relatively low
%number of radial modes, and we adopt. 
%and for \nuac\ we adopt the calculation laid out in Eq.
%(\ref{nuaciso}). 

\begin{figure}
\includegraphics[width=84mm]{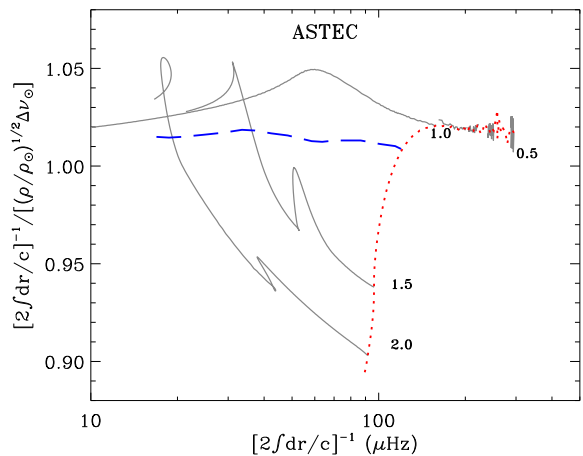}
%\epsscale{1.0}
%\plotone{f3.eps}
\caption{Ratio of \dnu\ between model calculations and solar scaling for \astec\
  grid ($Z=0.014$).  Blue dashed curve shows models with \teff\ $=6400\,$K.
  Annotation follows that of Fig.~\ref{f8}.  
 \label{f3}}  
\end{figure}

Finally, it is interesting to investigate how well the scaling relations for  
\dnu\ and \nuac\ (Eqs.~\ref{dnusc} and~\ref{numaxsc}) agree with the
model calculations (Eqs.~\ref{dnuint} and~\ref{nuaciso}). 
Figure~\ref{f3} examines how well $[2\int dr/c]^{-1}$ scales
with $\sqrt{\rho}$.   For cool models the agreement
is particularly good, almost independent of the evolutionary state, while it 
deteriorates slightly for hot models (those below the blue
line).  However, we note that $[2\int dr/c]^{-1}$ systematically
overestimates the large frequency separation for cool models, corresponding
to 2--3 per cent for the Sun. 
%We should also keep in mind that theoretical models
%of the Sun overestimate \dnu\ by 1\muhz\ due to incorrect modelling of the
%near-surface layers \citep{Dziembowski88,Dalsgaard88} and a similar effect
%presumably occurs for models of other stars \citet{Kjeldsen08a}.
%The maximum deviation is 10--15 per cent,
%corresponding to the hottest models in our grid with \teff$\sim10,000\,$K.   

\begin{figure}
\includegraphics[width=84mm]{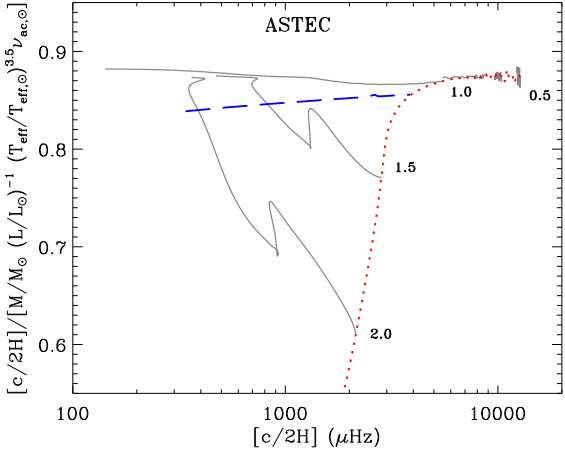}
%\epsscale{1.0}
%\plotone{f4.eps}
\caption{Ratio of \nuac\ between model calculations and scaling for \astec\
  grid ($Z=0.014$).  Annotation follows that of Fig.~\ref{f3}.  
\label{f4}} 
\end{figure}

In a similar way Fig.~\ref{f4} shows that the model calculations of
$c/2H$ follow the $M$\teff$^{3.5}/L$ scaling from the Sun quite well. The
ratio is always below unity because we evaluate $c/2H$ at 
the surface defined as $T=$ \teff, where $c/2H$ is slightly below its maximum
value, which occurs further out in the atmosphere. %[EVALUATING IT AT THE VERY END OF THE
%  ATM WHERE THE MODEL STOPS WOULD MAKE THE RATIO 1 FOR COOL STARS, BUT IT
%  WILL NOT BE SMOOTH BUT LOOK BAD IN A PLOT DUE TO NUMERICAL REASONS]
%For this plot we adopted \dnu$_0$ from Model S of the Sun \citep{Dalsgaard96}.   
As with \dnu, we conclude that the best agreement is for the cool
models.  We find similar results for the \yrec\ models.
%[IS THIS DETERIORATION FOR HOT STARS BECAUSE \teff\ IS NO LONGER A GOOD
%  APPROXIMATION TO THE MEAN LOCAL TEMPERATURE...IN ANY CASE WHAT IS THE
%  REASON WHY THAT SETS IN AT ABOUT 6400K..CAN ANYONE COMMENT ON THAT?]
%(\teff$\lesssim6200\,$K) 
%, with a maximum deviation of 20--30 per cent for the 
%hottest models in our grid.

\section{Conclusions}\label{conclusions}
This Letter points out that the
ratio \dnu/$\nu_{\mathrm{max}}^{0.77}$ varies very weakly with stellar
parameters, so that it is essentially constant.
Hence, if either one of 
these parameters is measured, this gives a very useful and robust estimate
of the other without any prior knowledge of the stellar global parameters,
$L$, $M$, \teff, and $Z$.  We anticipate this relation can be used to establish
the most plausible large separation in case of ambiguity for datasets where
\numax\ can be determined.  This has already been implemented
for automated analysis of \textit{Kepler} data
\citep{Hekker09a,Huber09,Mathur09}.  In addition, we showed that the well-used scaling
relations for \dnu\ and \numax\ agree within a few per cent with stellar
model calculations for cool models (\teff\ $\lesssim6400\,$K), from the main
sequence to the red giant branch, with a slightly increasing deviation for
hotter models.

\section*{Acknowledgments}
DS acknowledge support from the Australian Research Council.
WJC and YE acknowledge the support of the UK
Science and Technology Facilities Council (STFC).
DS would like to thank Douglas Gough for fruitful discussions.

%\clearpage

\bibliography{bib_complete}

%\clearpage
\label{lastpage}

\end{document}